\documentstyle[prb,aps,epsfig,amsfonts]{revtex}

\begin{document}

\twocolumn \psfull \draft

\wideabs{
\title{Which Kubo formula gives the exact conductance of a mesoscopic disordered system?}
\author{Branislav K. Nikoli\' c}
\address{Department of Physics, Georgetown University,
Washington, DC 20057-0995}

\maketitle

\begin{abstract}
In both research and textbook literature one often finds two
``different'' Kubo formulas for the zero-temperature conductance of a
non-interacting Fermi system. They contain a trace of the product
of velocity operators and single-particle (retarded and advanced)
Green operators: $\text{Tr} (\hat{v}_x \, \hat{G}^r \, \hat{v}_x
\, \hat{G}^a)$ or $\text{Tr} (\hat{v}_x \, \text{Im} \hat{G} \,
\hat{v}_x \text{Im} \, \hat{G})$. The study investigates the
relationship between these expressions, as well as the
requirements of current conservation, through exact evaluation of
such quantum-mechanical traces for a nanoscale (containing 1000
atoms) mesoscopic disordered conductor. The traces are computed
in the semiclassical regime (where disorder is weak) and, more
importantly, in the nonperturbative transport regime (including
the region around localization-delocalization transition) where
concept of mean free path ceases to exist. Since quantum
interference effects for such strong disorder are not amenable to
diagrammatic or nonlinear $\sigma$-model techniques, the
evolution of different Green function terms with disorder
strength provides novel insight into the development of an
Anderson localized phase.
\end{abstract}

}

\narrowtext

At first sight, the title of this paper might sound
perplexing. What else can be said about Kubo formula~\cite{kubo}
after almost a half of a (last) century of explorations in
practice, as well as through numerous re-derivations in both
research~\cite{baranger-kubo} and textbook~\cite{rammer,mahan}
literature? Kubo linear response theory (KLRT) represents the
first full quantum-mechanical transport formalism. It connects
irreversible processes in nonequilibrium to the thermal
fluctuations in equilibrium [fluctuation-dissipation theorem
(FDT)]. Therefore, the study of transport is limited to the
nonequilibrium states close to equilibrium. Nevertheless, the
computation of linear kinetic coefficients is greatly
facilitated since final expressions deal with equilibrium
expectation values  of relevant physical
quantities (which are much simpler than the corresponding
nonequilibrium ones~\cite{haug}). It originated~\cite{kubo_popular}
from the Einstein relation for the diffusion constant and
mobility of a particle performing a random walk.

Until the scaling theory of localization,~\cite{gang4} and ensuing
computation of the lowest order quantum correction, weak
localization~\cite{wl} (WL), to the Drude conductivity, it almost
appeared that microscopic and complicated Kubo formulation of
quantum transport merely served to justify the intuitive
Bloch-Boltzmann semiclassical approach~\cite{rammer} to
transport in weakly disordered ($k_F\ell \gg 1$, $k_F$ is the
Fermi wave vector and $\ell$ is the mean free path) conductors.
Furthermore, the advent of mesoscopic physics~\cite{lesh} has led
to reexamination of major transport ideas---in particular, we
learned how to apply properly KLRT to finite-size systems. Thus,
the equivalence was established~\cite{baranger-kubo} between the
rigorous Kubo formalism and heuristically founded Landauer-B\"
uttiker~\cite{lb} scattering approach to linear response
transport of non-interacting quasiparticles.~\cite{yudson}
This has emerged as an important tool in for studying mesoscopic
transport phenomena, where system size and interfaces through
which electron can enter or leave the conductor play an
essential role in determining the conductance.~\cite{landimry,andersonbook}

\begin{figure}
\centerline{\epsfig{file=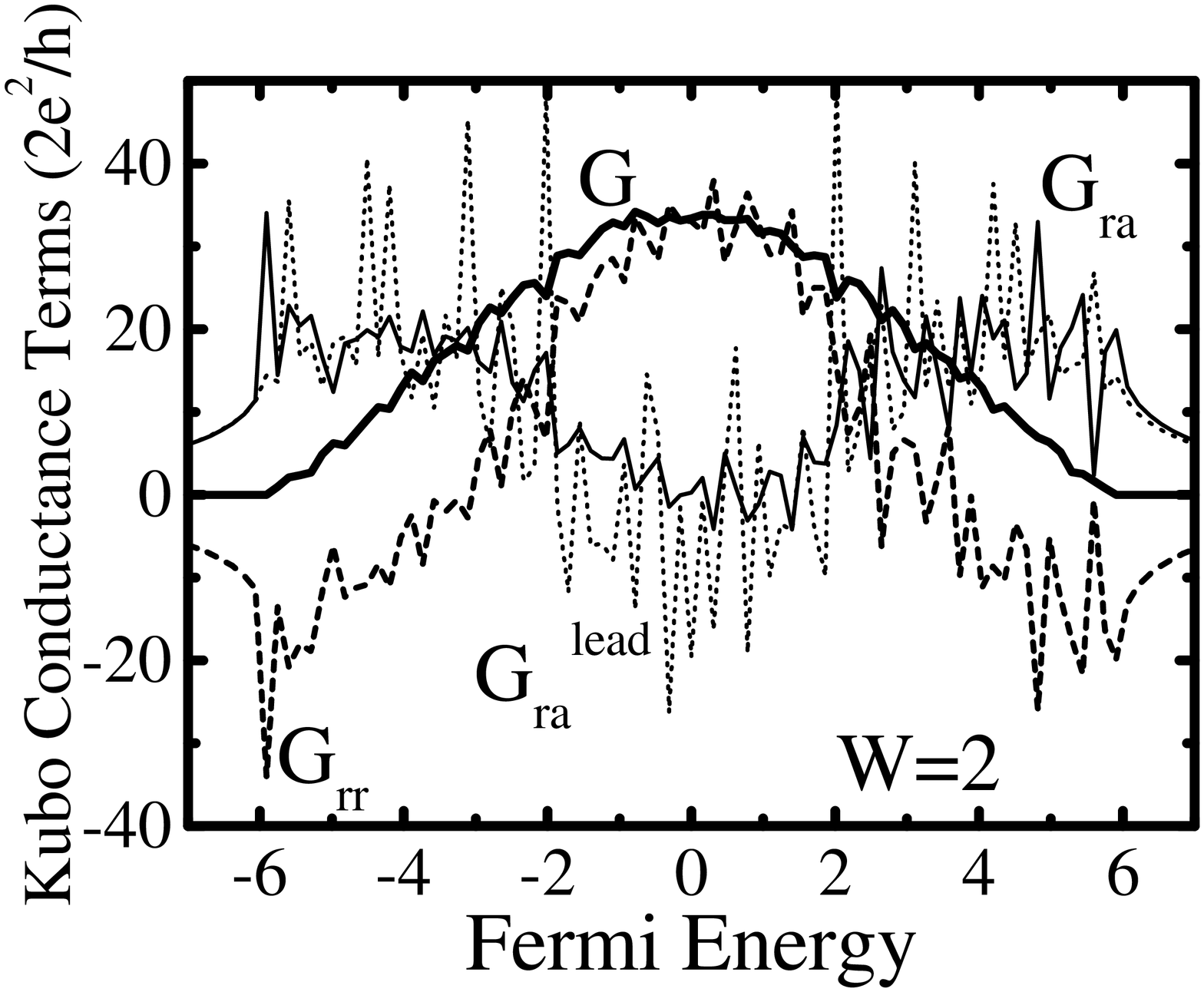,height=2.7in,angle=-90} }
\vspace{0.14in}
\centerline{\epsfig{file=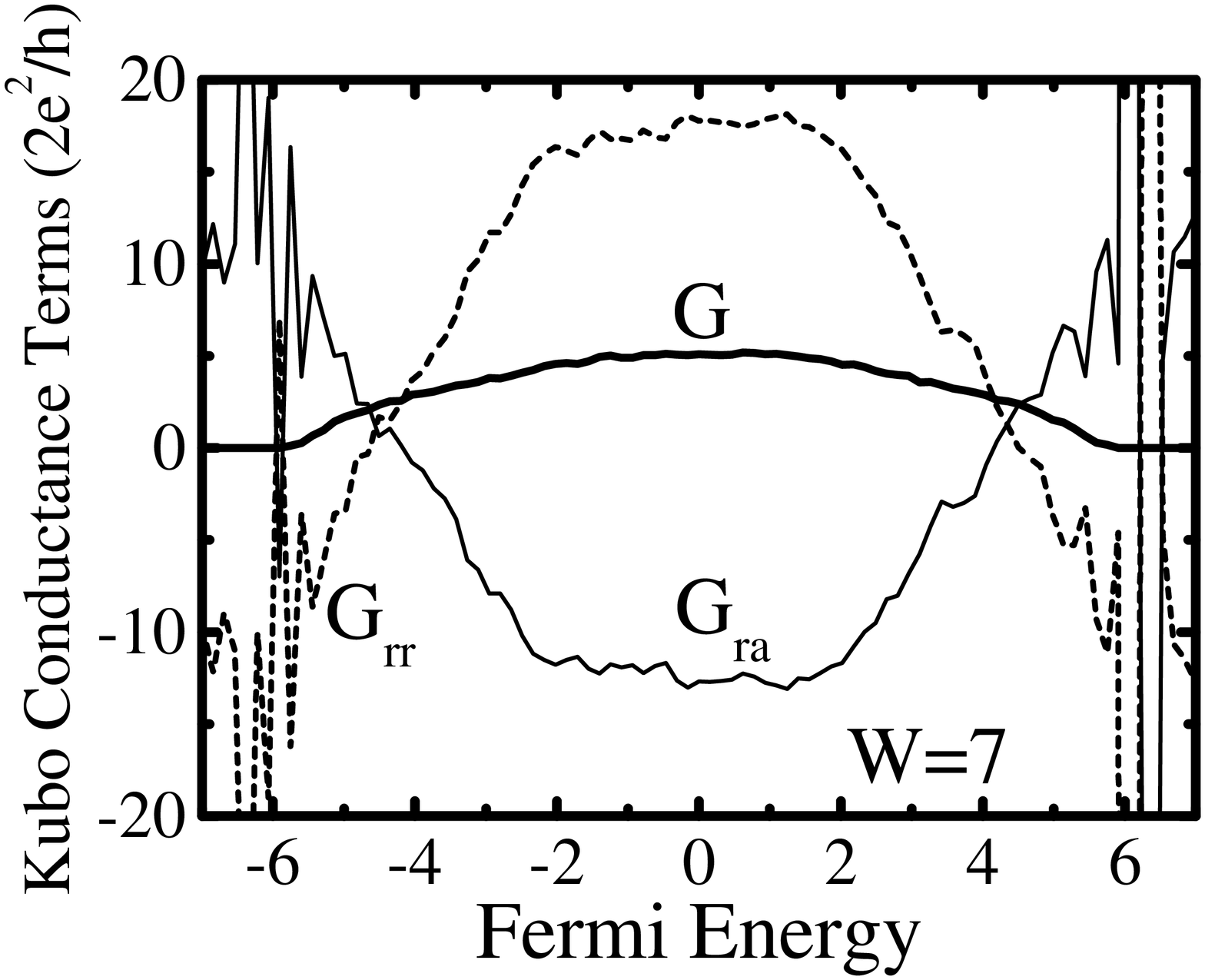,height=2.7in,angle=-90} }
\vspace{0.2in} \caption{Different terms in the Kubo formula for
the two-probe quantum conductance of a single finite-size sample
modeled on a simple cubic lattice $10^3$ by an Anderson model with
disorder strength $W=2$ [upper panel---single sample in the
semiclassical transport regime] or $W=7$ [lower
panel---disorder-averaged over 50 samples in the nonperturbative
transport regime $k_F \ell \protect \lesssim 1$]. The full Kubo
conductance (thick solid line) is given by the sum of terms
defined in Eq.~(\ref{eq:kubogra}) [thin solid line] and
Eq.~(\ref{eq:kubogrr}) [dashed line], $G=G_{ra}+G_{rr}$. The
respective traces in these expressions are performed only over
the states residing on the first two planes inside the sample.
The dotted line in the upper panel  represents $G_{ra}^{\rm lead}$
obtained by tracing over the two planes deep inside the left lead
(at the distance $10a$ away from the sample).}\label{fig:Wsurf}
\end{figure}

\begin{figure}
\centerline{\epsfig{file=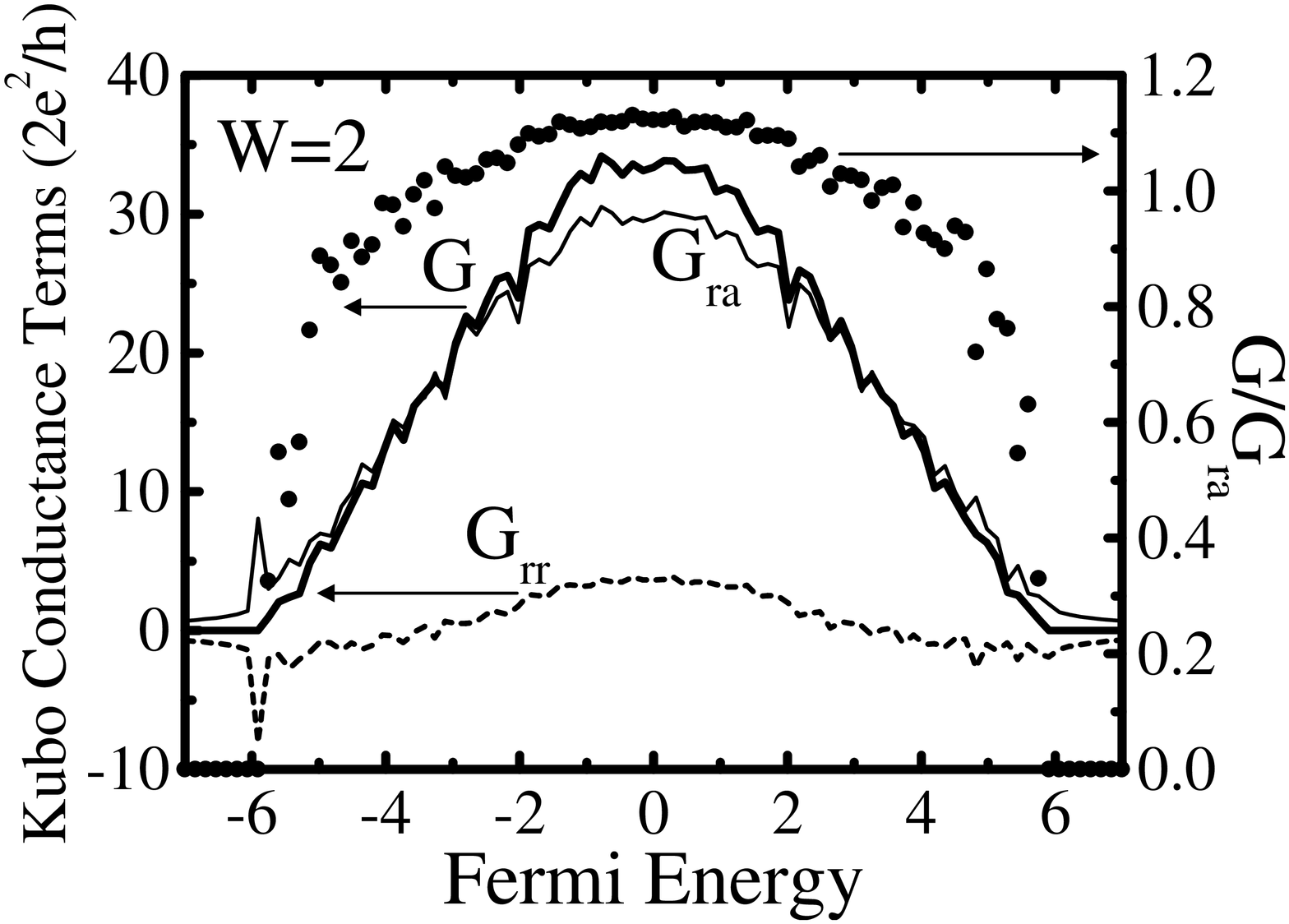,height=3.0in,angle=-90} }
\vspace{0.12in}
\centerline{\epsfig{file=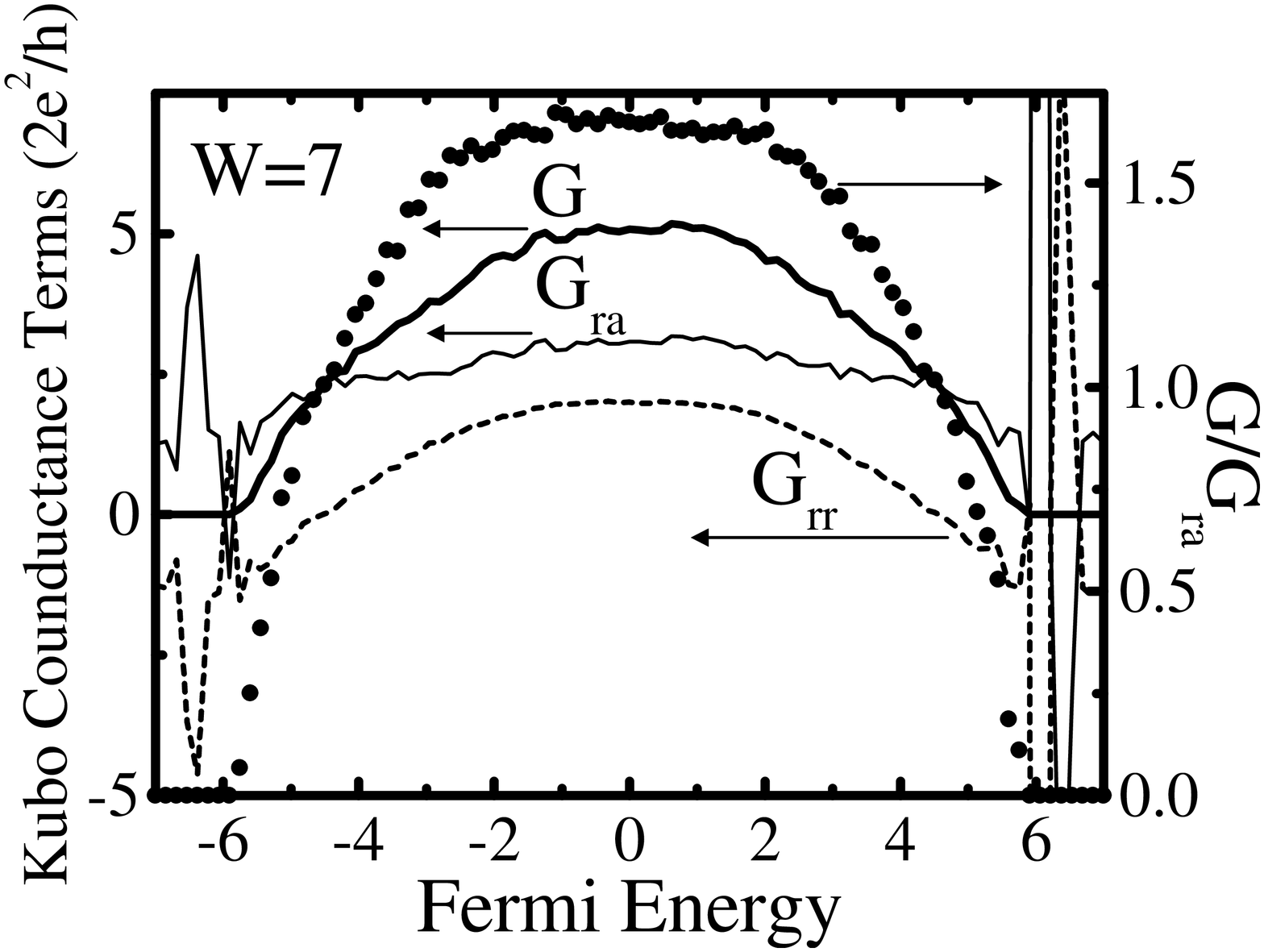,height=3.0in,angle=-90} }
\vspace{0.2in} \caption{Different terms in the Kubo formula
$G=G_{ra}+G_{rr}$ for the two-probe quantum conductance of the
same finite-size conductors as in Fig.~\ref{fig:Wsurf}, but with
respective traces in these expressions performed over the site
states inside the whole disordered sample.} \label{fig:Wfull}
\end{figure}

This study presents an exact evaluation of two different Kubo-type
expressions for the linear conductance of a mesoscopic disordered
conductor. Both expressions are frequently encountered in research as
well as textbook literature, and are displayed below. They
consist of a trace (or linear combination of such traces) over
the product of velocity operators $\hat{v}_x$ with retarded and
advanced single-particle Green operators
$\hat{G}^{r,a}=[E-\hat{H}\pm i0^+]^{-1}$, like $\text{Tr} \left
[\hat{v}_x\, \, \hat{G}^{r,a} \hat{v}_x \, \hat{G}^{r,a} \right
]$, where $\hat{H}$ is an equilibrium Hamiltonian (in the spirit
of FDT, it contains random and confining potentials, but not the
external electric field), and velocity operator is defined by $i
\hbar \hat{{\bf v}}=[\hat{{\bf r}},\hat{H}]$. These
quantum-mechanical traces are computed here, Figs.~\ref{fig:Wsurf}
and~\ref{fig:Wfull}, in the site representation (i.e., using
real-space Green functions) defined by a lattice model, such as
tight-binding Hamiltonian 
\begin{equation}\label{eq:tbh}
  \hat{H} = \sum_{\bf m} \varepsilon_{\bf m}|{\bf m} \rangle \langle {\bf m}|
  +  \sum_{\langle {\bf m},{\bf n} \rangle}
  t_{\bf mn} |{\bf m} \rangle \langle {\bf n}|.
\end{equation}
on a hypercubic lattice $N^d$ of size $L=Na$ ($a$ being the
lattice constant). Here $t_{\bf mn}$ is the nearest-neighbor
hopping integral between $s$-orbitals $\langle {\bf r}|{\bf m}
\rangle = \psi({\bf r}-{\bf m})$ on adjacent atoms located at
sites ${\bf m}$ of the lattice ($t_{\bf mn}=1$ inside the sample
defines the unit of energy). The disorder is simulated by taking
random on-site potential such that $\varepsilon_{\bf m}$ is
uniformly distributed over the interval $[-W/2,W/2]$, which is 
the so-called Anderson model of localization. I emphasize
the requirements of current conservation throughout this
analysis, which will allow us to understand the features of
different trace expressions introduced above.

\begin{figure}
\centerline{
\psfig{file=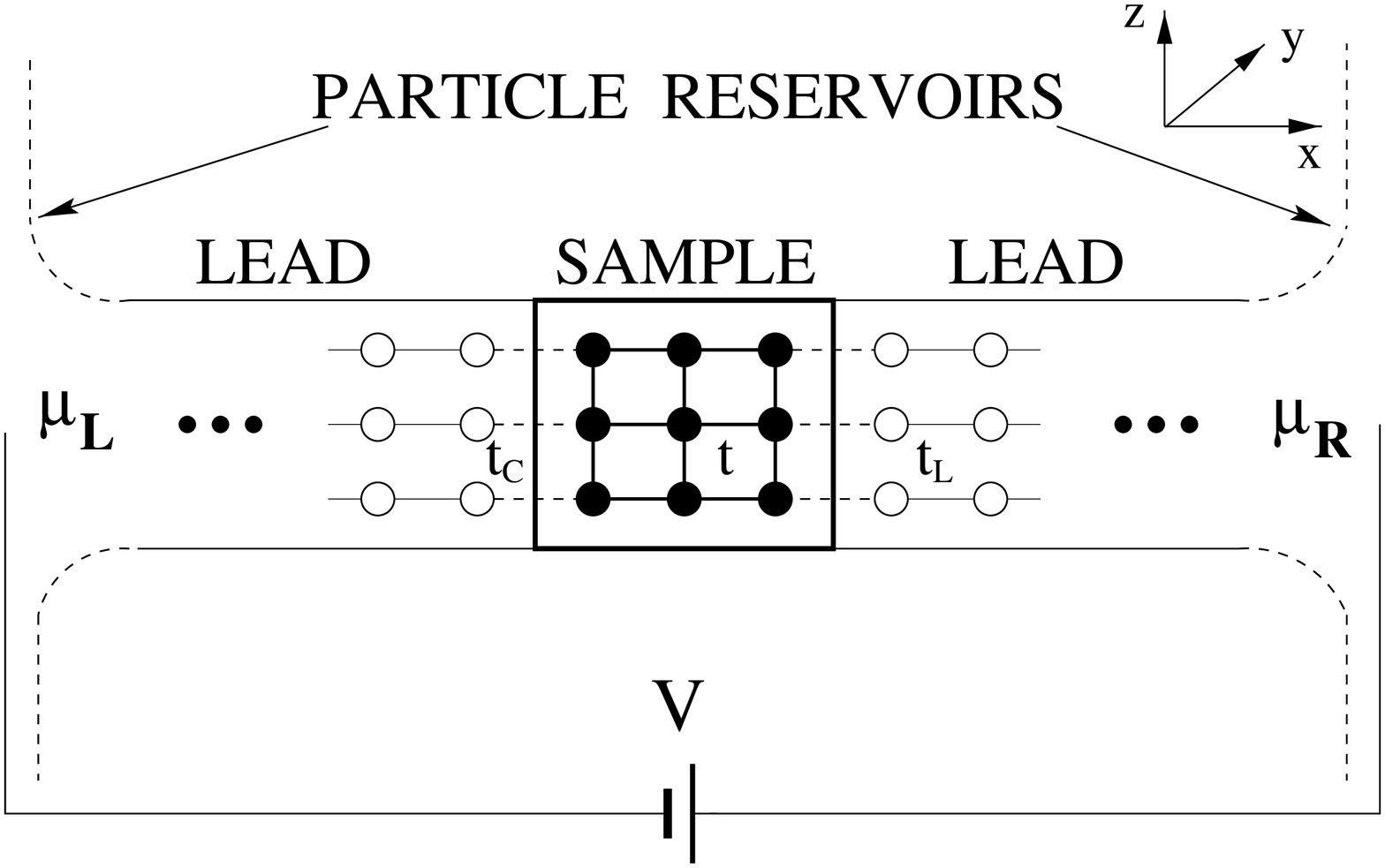,height=2.3in,width=3.0in,angle=0} }
\vspace{0.2in} \caption{A two-dimensional version of the actual 3D
model of a two-probe measuring geometry employed here. Each site
hosts a single $s$-orbital which hops to six (or fewer for surface
atoms) nearest neighbors. The hopping matrix element is $t$
(within the sample), $t_{\text L}$ (within the leads), and
$t_{\text C}$ (coupling of the sample to the leads). The leads are
semi-infinite and are considered to be connected at $\pm \infty$
to ``reservoirs'' biased by the potential difference
$eV=\mu_L-\mu_R$.} \label{fig:setup}
\end{figure}

The mesoscopic methods (mesoscopic Kubo~\cite{baranger-kubo,verges} 
or, equivalent, Landauer~\cite{lb} formula) make it possible to get 
the exact zero-temperature (i.e., quantum) conductance of a finite-size
sample attached to semi-infinite disorder-free leads. Although
KLRT is a standard formalism for introducing the many-body
physics into the computation
of transport coefficients,~\cite{mahan} here the focus is on the transport
properties determined by scattering of non-interacting
(quasi)electrons on impurities.  The ``old'' Kubo formula~\cite{kubo_volume}
for the macroscopic volume-averaged longitudinal DC conductivity
at zero temperature ($E \equiv E_F$ in all formulas below,
$E_F$ being the Fermi energy) of a non-interacting Fermi
gas described by a single-particle Hamiltonian $\hat{H}$ is
given by
\begin{equation}\label{eq:kubo}
   \sigma_{xx}  =  \frac{2\pi e^2 \hbar}{\Omega} \, \text{Tr} \left
   [\hat{v}_x\, \delta(E-\hat{H}) \, \hat{v}_x  \, \delta(E-\hat{H})  \right
  ],
\end{equation}
where factor of two accounts for the spin degeneracy. The Kubo conductivity
relates the spatially averaged current
 ${\bf j}=\int d{\bf r} \,
{\bf j}({\bf r})/\Omega$ to the spatially-averaged electric field
${\bf j}=\sigma {\bf E}$, where thermodynamic limit
$\Omega=L^d \rightarrow \infty$ (while keeping the impurity concentration
finite) is implied to get the unambiguous intensive
quantity~\cite{qt,gang4} (and well-defined
steady state). For electrons in a random potential further averaging
should be performed over the disorder to get $\sigma$ as a
material constant.~\cite{bastin} On the other hand, quantum corrections to the
conductivity are non-local~\cite{wl} on the scale of the
dephasing length $L_\phi \gg \ell$. This invalidates the concept
of local quantities, like conductivity, in mesoscopic samples,
which are smaller than $L_\phi$ and thereby effectively at
$T=0$. Therefore, mesoscopic transport has to be described in
terms of sample-specific quantities, like conductance, which
describe a given sample measured in a given
manner~\cite{baranger-kubo} (i.e., more generally, conductance
coefficients $I_p=\sum_q g_{pq} V_q$ in the Ohm's law for a
multi-probe geometry, where several leads are attached to the
sample to feed the current $I_p$ or measure the voltages $V_q$),
or alternatively, non-local conductivity tensor introduced below.
Switching to conductance leads to a following Kubo expression
\begin{mathletters}
\label{eq:kubog}
\begin{eqnarray}
 G & = & \frac{4e^2}{h} \frac{1}{L^2} \, \text{Tr} \left [\hbar
 \hat{v}_x \,
 \text{Im} \,  \hat{G} \, \hbar \hat{v}_x \, \text{Im} \, \hat{G} \right ], \\
 \text {Im} \, \hat{G} & = & \frac{1}{2i} (\hat{G}^r - \hat{G}^a) = -\pi
 \delta(E-\hat{H}).
\end{eqnarray}
\end{mathletters}
Here the definition of retarded ($r$) or advanced ($a$)
single-particle Green operator $\hat{G}^{r,a}=[E-\hat{H} \pm
i0^+]^{-1}$ requires a numerical trick to handle the small
imaginary part $i0^+$, which then spoils the prospect of
obtaining the exact zero-temperature
conductance.~\cite{thouless,krey} Once the semi-infinite clean
leads are attached to the finite sample (at planes $1$ and $N$
along $x$-axis for a two-probe geometry, Fig.~\ref{fig:setup}),
the ``self-energy''
$\hat{\Sigma}^{r,a}=\hat{\Sigma}_L^{r,a}+\hat{\Sigma}_R^{r,a}$,
arising from the ``interaction'' with the leads ($L$-left,
$R$-right), provides a well-defined imaginary part in the
definition of the Green operators~\cite{datta}
\begin{equation}\label{eq:greenop}
\hat{G}^{r,a}=[E-\hat{H}-\hat{\Sigma}^{r,a}]^{-1}.
\end{equation}
The Green function $\hat{G}^{r,a}({\bf n},{\bf m})=\langle {\bf
n} | \hat{G}^{r,a} | {\bf m} \rangle$ describes the propagation of
electron between two sites inside an open conductor ($L_\phi=L$
in the two-probe geometry). The self-energy terms are
given~\cite{datta,nikolic_qpc} by $\hat{\Sigma}^{r}_{L,R}({\bf
n},{\bf m}) = (t_{\rm C}^2/2t_{\rm L}^2) \, \hat{g}_{L,R}^r({\bf
n}_S,{\bf m}_S)$ with $\hat{g}_{L,R}^r({\bf n}_S,{\bf m}_S)$
being the surface Green function of the bare semi-infinite lead
between the sites ${\bf n}_S$ and ${\bf m}_S$ located on the end
atomic layer of the lead (and adjacent to the corresponding sites
${\bf n}$ and ${\bf m}$ inside the conductor). It has an
imaginary part only for $|E| \le 6t_{\rm L}$, which means that
$G(E_F)$ goes to zero at band edge of a clean lead $|E_F|=6t_{\rm
L}$ because there are no states in the leads beyond this energy
which can carry the current. Here the leads will be described by
the TBH with $\varepsilon_{\bf m}=0$ and $t_{\bf mn}=t_{\rm L}$; the
hopping between the sites in the lead and the sample is $t_{\bf mn}=t_{\rm C}$, 
as illustrated in Fig.~\ref{fig:setup}.

I use the term ``mesoscopic Kubo formula'' for
Eq.~(\ref{eq:kubog}) with the Green operators~(\ref{eq:greenop})
plugged in. This formula is exactly
equivalent~\cite{baranger-kubo} to a two-probe Landauer
formula~\cite{lb} for the conductance measured between two points
deep inside macroscopic reservoirs to which the leads are
attached at infinity.~\cite{landvskubo} Thus, it is conceptually
different from the ``plain'' Kubo formula~(\ref{eq:kubo}) which
follows from combining the conductance of smaller parts $G =
\sigma L^{d-1}/L$, thereby implying local description of transport
which becomes applicable only at sufficiently high temperatures.
Such mesoscopic formulas provide means to compute the quantum conductance---a
sample specific quantity which takes into account the finite
system size, measuring geometry, arrangement of impurities,
non-local features of quantum transport, and can describe
ballistic transport (where local relation ${\bf j}=\sigma {\bf
E}$ does not hold). Attempts to use the original Kubo formulas on
finite samples, throughout  premesoscopic
history~\cite{thouless,krey} of the Anderson localization theory,
were thwarted with ambiguities, which can be traced back to the
general questions on the origin of dissipation.~\cite{landauer_ph}
This stems from the fact that stationary regime cannot be reached
unless the system is infinite or coupled to a
thermostat. From the practical
point of view, leads make the system infinite by opening
the sample, and therefore eliminating the technical obstacles in
handling of the discrete spectrum of finite samples.~\cite{closed}
Furthermore, the usage of semi-infinite leads allows us to bypass
explicit modeling of reservoirs in the computation of conductance
since ``hot'' electrons which escape into the leads (due to the
broadening of energy levels) do not come back in a phase-coherent
fashion. The concept of reservoirs was always essential part of
Landauer's subtle arguments.~\cite{landauer_ph} They provide
dissipation and therefore the steady state.~\cite{buttler_footnote}
However, the computation of conductance, as a measure the dissipation,
involves only a conservative Hamiltonian of noninteracting
electrons scattered on impurities (i.e., when computing
linear conductance of a mesoscopic system one usually
does not deal explicitly with electron-electron and
electron-phonon interactions~\cite{yudson}).

The conductance computed from the mesoscopic quantum expression is
exact, but characterizes the whole {\em sample+leads} system in
the spirit of quantum measurement theory since leads can also be
considered as the ``macroscopic measuring
apparatus''.~\cite{efetovbook,hans} However, for disordered enough
sample (and not too narrow leads or too small $t_{\rm C}$) the
conductance is determined mostly by the disordered region
itself.~\cite{braun,nikolic_qpc} This exactness makes it possible
to compute the transport properties in both semiclassical regime
(where Boltzmann theory and perturbative quantum corrections are
applicable since $k_F\ell \gg 1$), as well as in the
nonperturbative transport regime~\cite{nikolic_qrho} where
semiclassical concepts, like $\ell$, loose
their meaning. Although the distinction between $k_F\ell \sim 1$
(where semiclassical theory, including the perturbative quantum
corrections, breaks down) and the criterion $G \sim 2e^2/h$ for
the localization-delocalization (LD) critical point has been
known since the scaling theory of localization,~\cite{gang4} it
is not uncommon practice to find these two different boundaries
of transport regimes confused. As the disorder is increased, a
sample goes from the semiclassical transport regime, through a
vast region between $k_F\ell \sim 1$ and LD transition (e.g., $6
\lesssim W \lesssim 16.5$ at $E_F=0$ in the Anderson
model~\cite{nikolic_qrho}), and finally enters into a localized
phase. Since the nonperturbative transport regime lacks small
parameter required by present analytical schemes, the numerical
techniques employed here are the only way to gain insight into
the quantum effects beyond the lowest order corrections, like WL,
or resummation of all such perturbative quantum correction within
the nonlinear $\sigma$-model formalism.~\cite{wegner,lerner}

Another expression is often encountered in the
literature~\cite{rammer,qt,haug,datta,nazarov} for both
conductance of finite-size systems and conductivity of infinite
systems. It gives the conductance through an apparently different
trace, $G \propto \text{Tr} \left [\hbar \hat{v}_x \, \hat{G}^r
\, \hbar \hat{v}_x \, \hat{G}^a \right ]$. While some textbooks
quote this only as a convenient approximation to the
disorder-averaged full formula,~\cite{rammer,qt,haug} sometimes
the claim goes further to say that such trace is equivalent to
the Landauer formula, and moreover it can be evaluated at any
cross section inside the disordered region, thus relaxing the
requirement of perfect leads ``as the weakest point of the
Landauer formalism''.~\cite{nazarov} Namely, $\hat{G}^r
\hat{G}^r$ or $\hat{G}^a \hat{G}^a$ terms can be reduced to a
single Green function via a Ward identity in weakly disordered
conductors,~\cite{abrikosov,rammer} and are therefore related to
the density of states. They are ``abandoned'' in the limit
$k_F\ell \gg 1$ since they do not generate interesting
contributions to WL or mesoscopic fluctuations.~\cite{qt} In fact,
both the diffusion modes of the Diffuson-Cooperon diagrammatic
perturbation theory and different versions of the nonlinear
$\sigma$-model (NLSM) are derived~\cite{efetovbook,lerner} by
considering only the term containing the product $\hat{G}^r
\hat{G}^a$. The NLSM is a quantum field theory~\cite{efetovbook}
of weakly disordered mesoscopic conductors where diffusion modes
of the diagrammatic perturbation theory play the role of soft
modes responsible for long-range spatial correlations of local 
density of states, mesoscopic  fluctuations of global quantities, 
and nonlocal corrections to  conductivity.~\cite{simons} It makes 
it possible to handle the  breakdown of perturbation theory due to 
the proliferation of such modes~\cite{simons} by summarizing all WL-type 
corrections to conductivity.~\cite{wegner} This then justifies the 
phenomenological  one-parameter scaling theory~\cite{gang4} (if not 
rigorously,  then at least qualitatively), and explains the LD 
transition in  $2+\epsilon$ dimensions where Anderson localization 
occurs at weak disorder $k_F \ell \gg 1$.

To remove possible confusion,~\cite{efetovbook} it should be
emphasized that conductance coefficients $g_{pq}$, obtained by
integrating Kubo non-local conductivity tensor $\mbox{\b{$\sigma$}}
({\bf r},{\bf r}^\prime)$ over the cross sections in the leads $p$, $q$
\begin{equation}\label{eq:cond}
  g_{pq}= - \int\limits_{S_p} \!  \int\limits_{S_q} d{\bf S}_p \cdot \mbox{\b{$\sigma$}}
  ({\bf r},{\bf r}') \cdot d{\bf S}_q,
\end{equation}
also contain ${\rm Im} \, \hat{G}$ for $p=q$, while for $p \neq q$ only
the terms involving $\hat{G}^r \hat{G}^a$ are non-zero.~\cite{baranger-kubo}
The cross sections $S_p$ and $S_q$ are to be chosen far enough
from the sample, where all evanescent modes have died out. This
not only provides the rigorous foundation for the Landauer
formalism, but also clarifies some subtle points in the Kubo
formalism (like disorder averaging~\cite{gijs} and independence of
linear transport properties from the nonequilibrium charge
redistribution~\cite{kane}). By writing the full Kubo
formula~(\ref{eq:kubog}) as a sum $G = G_{ra} + G_{rr}$, I
introduce the ``Kubo conductance terms''
\begin{eqnarray}
 \label{eq:kubogra}
 G_{ra} & = & \frac{2e^2}{h} \frac{1}{L^2} \, \text{Tr} \left [\hbar \hat{v}_x
  \,  \hat{G}^r \, \hbar \hat{v}_x \, \hat{G}^a \right ],  \\
  \label{eq:kubogrr}
  G_{rr} & = & - \frac{2e^2}{h} \frac{1}{L^2} \, \text{Tr} \left [\hbar \hat{v}_x
  \,  \hat{G}^r \, \hbar \hat{v}_x \, \hat{G}^r \right ].
\end{eqnarray}
Obviously, if in some transport regime conductance can be
obtained from the trace in $G_{ra}$, the other term $G_{rr}$
has to vanish, at least approximately.

Before embarking on the direct evaluation of these expressions
for a conductor described by TBH~(\ref{eq:tbh}), the crucial
point is to understand the way (i.e., space of states $| {\bf m}
\rangle$) in which the traces should be performed, ${\rm
Tr}(\ldots) = \sum_{\bf m} \langle {\bf m}|(\ldots)|\bf m\rangle$.
Na\"{\i}vely, in the site representation it would appear that
trace in formula~(\ref{eq:kubog}) should include site states
inside the whole sample. However, once the current conservation $
\nabla \cdot {\bf j}({\bf r}) = 0$ is invoked this becomes
extraneous. All Kubo conductivity or conductance formulas stem
from the more fundamental quantity in KLRT, the non-local
conductivity tensor relating local current density to the local
electric field,
\begin{equation}\label{eq:nlct}
{\bf j}(\bf{r}) = \int \! d {\bf r}^\prime \, \mbox{\b{$\sigma$}}
({\bf r},{\bf r}^\prime) \cdot {\bf E}({\bf r}^\prime)
\end{equation}
This tensor is obtained as a response to an external field only since
corrections to the current due to the field of induced charges go
beyond the linear transport regime~\cite{stone_book,nikolic_cpc}
(i.e., one does not have to engage in a much more cumbersome task of
finding the response to a full electric field inside the
conductor~\cite{mahan}). In application of KLRT to mesoscopic
systems, $\mbox{\b{$\sigma$}} ({\bf r},{\bf r}^\prime)$ is a
sample-specific quantity, i.e., defined for each impurity configuration
and arrangement of the leads~\cite{baranger-kubo}, and is in fact
non-local even after disorder averaging.~\cite{kane} Because it is
not directly measurable, some volume averaging is needed to get
the quantities that can be related to experiments
\begin{equation}\label{eq:volumecond}
  G =  \frac{1}{V^2} \int\limits_{\Omega}  d{\bf r} \, d{\bf r}' \, {\bf E}({\bf r})
  \cdot \mbox{\b{$\sigma$}} ({\bf r},{\bf r}') \cdot {\bf E}({\bf
  r}^\prime).
\end{equation}
Here $V$ is  the bias voltage ($V \rightarrow 0$ in the linear
transport regime), e.g., in the case of two-probe geometry
$eV=\mu_L-\mu_R$, where leads are in equilibrium with two
\begin{figure}
\centerline{\epsfig{file=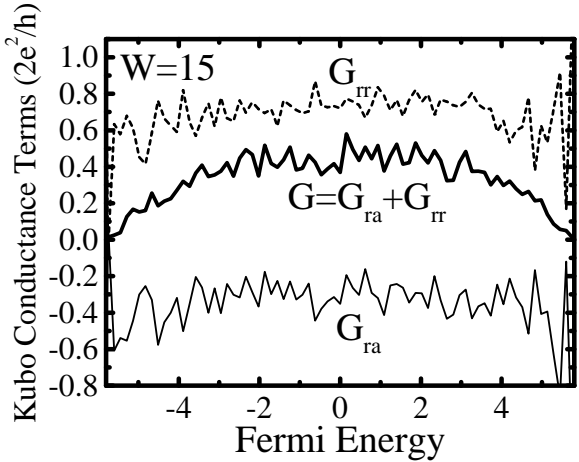,height=3.0in,angle=-90} }
\vspace{0.2in} \caption{Different terms in the Kubo formula
$G=G_{ra}+G_{rr}$ for the two-probe quantum conductance of the
finite-size conductor modeled on a lattice $10^3$ with diagonal
disorder $W=15$ (whole band becomes localized at $W_c \simeq
16.5$). Respective traces in these expressions are performed over
the site states inside the whole disordered sample and
disorder-averaged over 50 realizations.} \label{fig:W15full}
\end{figure}
macroscopic reservoirs characterized by a constant chemical
potentials $\mu_L$ and $\mu_R$ (Fig.~\ref{fig:setup}). Although
this expression contains the local electric field ${\bf E}({\bf
r}) = \nabla \mu({\bf r})/e$ inside the conductor, because of
current conservation entailing $ \nabla \cdot \mbox{\b{$\sigma$}}
({\bf r},{\bf r}')=\mbox{\b{$\sigma$}} ({\bf r},{\bf r}') \cdot
\nabla'=0$ (which is a special case, in the absence of magnetic
field, of a more general theorem~\cite{baranger-kubo}), the
conductance will not depend on this field factors~\cite{kane}. In
the case of TBH~(\ref{eq:tbh}) with nearest-neighbor hopping the
expectation value of the velocity operator in the site
representation $\langle {\bf m} |\hat{v}_x| {\bf n} \rangle =
(i/\hbar)\, t_{\bf  mn} \left( m_x - n_x \right)$ is
 non-zero only between the states residing on adjacent planes
(technical details of a route from Eq.~(\ref{eq:volumecond}) to the
 trace involving velocity operator are meticulously covered in
Ref.~\onlinecite{baranger-kubo}). Thus, the minimal space choice here
is defined by taking ${\bf E}({\bf r})$ to be non-zero on two
adjacent planes, while the standard textbook assumption of a
homogeneous field~\cite{rammer,kubo_volume}
$|{\bf E}|=V/L$ throughout the sample leads to Eqs.~(\ref{eq:kubo}),
({\ref{eq:kubog}). Blind tracing over the whole sample would
give simply the conductance multiplied by the square of the number
of pairs of adjacent planes, meaning that such trace should
be divided by $(N-1)^2a^2$ to get $G$ [when trace is performed
only over the arbitrary two adjacent planes, $L^2$ in Eq.~({\ref{eq:kubog})
is replaced by $a^2$]. The physical content of this statement is
simple: current $I=GV$ is the same on each cross section.
Therefore, the conductance depends only on the total voltage drop
over the sample, and not on the local current density and
electric-field distributions.

At this point, it is worth mentioning that another expression is
frequently employed in the real-space computational practice. It
stems from the linear response limit of a formula derived by the
Keldysh technique (for non-interacting~\cite{caroli} or
interacting systems~\cite{meir})
\begin{figure}
\centerline{\epsfig{file=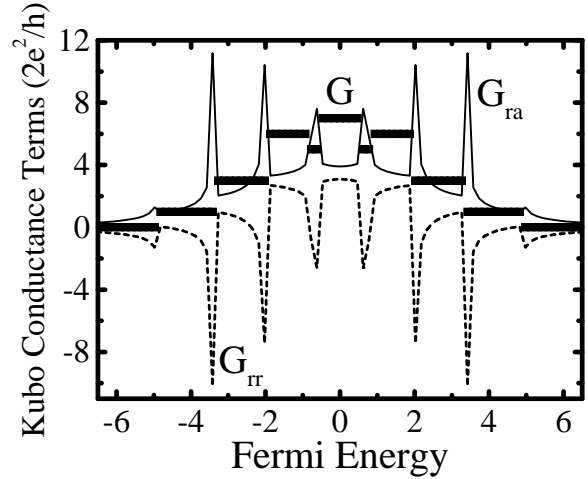,height=3.0in,angle=-90} }
\vspace{0.2in} \caption{Different terms in the Kubo formula
$G=G_{ra}+G_{rr}$ for the two-probe quantum conductance of a clean
finite-size conductor modeled on a lattice $3^3$
($\varepsilon_{\bf m}=0$, $t=t_{\rm L}=t_{\rm C}=1$). The
conductance as a function of the Fermi energy changes in steps
corresponding to the number of open conducting channels (defined
by 9 quantized transverse propagating modes).} \label{fig:cq}
\end{figure}
\begin{equation}\label{eq:greenlandauer}
  G  =  \frac{4e^2}{\pi \hbar} \, \text{Tr} \left ( \text{Im} \, \hat{\Sigma}_L \,
  \hat{G}^{r}_{1 N} \,
  \text{Im} \, \hat{\Sigma}_R \, \hat{G}^{a}_{N 1} \right ).
\end{equation}
Here $\hat{G}^{r}_{1 N}$, $\hat{G}^{a}_{N 1}$ are the submatrices
of the full Green function $\hat{G}^{r,a}({\bf m},{\bf n})$ whose
elements connect the layer $1$ to layer $N$ of the sample. Therefore, 
only a block $N^2 \times N^2$ of the whole $N^3 \times N^3$ matrix 
$\hat{G}^r({\bf n},{\bf m})$ is needed to compute the conductance. 
Although such ``partial'' knowledge of the whole Green function and the trace over
matrices of size $N^2$ in Eq.~(\ref{eq:greenlandauer}) are different
from the corresponding counterparts required in application of the
 mesoscopic Kubo formula (where minimal trace goes over $2N^2 \times 2N^2$
 matrices), the final result for the conductance is the same. Positive
definiteness of the operators $-2\, \text{Im} \,
\hat{\Sigma}_{L,R}$ makes it possible to find their square root
and recast the expression under the trace of
Eq.~(\ref{eq:greenlandauer}) as a Hermitian operator. The
expression~(\ref{eq:greenlandauer}) then looks like the two-probe
Landauer formula involving a transmission matrix ${\bf t}$
\begin{eqnarray}\label{eq:ttlandauer}
  G & = & \frac{e^2}{\pi \hbar} \, \text{Tr} \, ({\bf t t}^{\dag}) =
  \frac{e^2}{\pi \hbar} \sum_{n=1}^{N^2} T_n, \\
  {\bf t} & = & 2 \sqrt{-\text{Im} \, \hat{\Sigma}_L} \, \hat{G}^{r}_{1 N}
  \sqrt{-\text{Im}\, \hat{\Sigma}_R},
  \label{eq:t}
\end{eqnarray}
or transmission eigenvalues $T_n$ when the trace is evaluated in a
basis which diagonalizes ${\bf t t}^{\dag}$.
Moreover, it is more efficient from a practical point of view
since Green function techniques used to evaluate
Eq.~(\ref{eq:greenlandauer}) do not require to know exact
asymptotic eigenstates in the leads (as is the case with
``original'' Landauer formulation relying on the knowledge of the
scattering basis of wave functions defined within the asymptotic
regions of the leads). This becomes {\em sine qua non} when
computing the conductance of systems with complicated ``leads'',
e.g., like in the case of atomic-size point contacts.~\cite{yeyati}

The trace expressions like~(\ref{eq:kubogra})
and~(\ref{eq:kubogrr}) {\em do not} conserve the current. Therefore,
the suggested strategy is to evaluate $G_{ra}$ and $G_{rr}$ by
tracing over the states located on two planes inside the sample
(or inside the leads), as well as over the whole sample (and see
how close $G_{ra}$ can get to $G$). Different types of conductors
are chosen for this evaluation: weakly disordered with $W=2$
(e.g., $\ell \approx 9a$ at $E_F=0$), and strongly disordered
conductors with $W=7$ (for which unwarranted use of the
Drude-Boltzmann formula would give~\cite{nikolic_qrho} $\ell <
a$). The exact computation of these traces is shown in the
upper panels of Figs.~\ref{fig:Wsurf} and \ref{fig:Wfull} for a
single impurity configuration of $W=2$ (here $G_{ra}$ is also
computed by tracing over two planes deep inside the leads, which
corresponds to integrating $\mbox{\b{$\sigma$}} ({\bf r},{\bf
r}^\prime)$ over such cross section, as discussed above). The
lower panels of Figs.~\ref{fig:Wsurf}, \ref{fig:Wfull} plot
disorder-averaged quantities over an ensemble of impurity
configuration for disorder strength $W=7$. In both ways of
tracing, the sum of two terms $G_{ra}$ and $G_{rr}$ gives the
full expression for conductance $G(E_F)$, which have to cancel
each other for $|E_F|>6t$ in order to ensure vanishing of
$G(E_F)$ when $t_{\rm L}=t$ is chosen.

The result for $W=7$ disorder belongs to the nonperturbative
transport regime. It is interesting, therefore, to follow the
behavior of $G_{ra}$ and $G_{rr}$ terms further throughout this
regime, eventually reaching $W_c \simeq 16.5$ where the whole band
becomes localized.~\cite{slevin} The generic LD transition point
in three-dimensions (3D) is beyond NLSM treatment inasmuch as it
corresponds to a strong coupling limit in the field theoretical
language. Furthermore, recent description~\cite{vlada} of Anderson
localization in terms of an order parameter, obtained from a
theory based on local approximation, suggest that probing the
instability of the delocalized phase through WL-type corrections
might be a daunting road to reach the LD transition in 3D.
Figure~\ref{fig:W15full} shows that around $W_c$ the conductance
$G \sim 2e^2/h$ is mostly defined by the $G_{rr}$ term---a
situation completely opposite to the weak disorder finding ($W=2$
in Figs.~\ref{fig:Wsurf} and~\ref{fig:Wfull}). Therefore, at some
intermediate disorder $7 < W < 10$, within the nonperturbative
transport regime, a transition occurs from $G_{ra} > G_{rr}$ to
$G_{rr} > G_{ra}$, ending up eventually with a case $G_{ra} < 0 <
G_{rr}$ around the LD critical point.

An inquisitive reader might have come up by now with the question as
to what happens in the clean case. A mesoscopic ballistic sample
attached to two leads has non-zero point contact
conductance~\cite{nikolic_cpc} of a purely geometrical origin
since leads are widening into macroscopic reservoirs at infinity,
where reflection occurs when the large number of conducting
channels in the macroscopic reservoir matches the small number of
channels in the lead.~\cite{imry} Such point contact conductance
is quantized,~\cite{qt} as a function of sample
width~\cite{datta} or Fermi energy,~\cite{nikolic_qpc} which
becomes conspicuous when the number of quantized transverse
propagating modes (i.e., the sample cross section) is small
enough. A clean ($\varepsilon_{\bf m}=0$) toy sample $3 \times 3
\times 3$ illustrates conductance quantization in
Fig.~\ref{fig:cq}, as obtained from the two-probe Kubo formula
for $G$. In this case, both terms $G_{ra}$ and $G_{rr}$ are
contributing in a non-trivial fashion to a stepwise conductance.

What can be learned from these numbers is that
$G_{ra}$~(\ref{eq:kubogra}) can serve as a decent {\em
approximation} to the exact Kubo formula for the quantum
conductance $G$ only in very weakly disordered conductors.
However, because of not conserving the current, the trace in
$G_{ra}$ has to be performed over the whole disordered region
(which is an enormous computational effort, and is therefore
useless in the real-space computational practice). The essential
outcome of this exercise is the explicit quantification of the
difference between $G$ and $G_{ra}$. This, together with current
conservation, are important points to bear in mind when using
simplified expression $G_{ra}$ in analytical derivations and
arguments of the quantum transport theory.~\cite{rammer,qt,datta}
Finally, the strong disorder (nonperturbative) behavior of two
different Green function terms, comprising the Kubo formula for
quantum conductance, might provide a clue for the inadequacy of
attempts to analyze genuine Anderson transition in 3D by probing
instability of metallic phase to weak localization (perturbative)
corrections.

I am grateful to P. B. Allen for initiating my interest into
intricacies of the Kubo formalism, and to I. L. Aleiner, S.
Datta, J. K. Freericks, A. Maassen van den Brink, and J. A. Verg\'
es for valuable discussions. Special thanks goes to three unknown
Physical Review referees for their interest into improving the
manuscript. Financial support from ONR grant N00014-99-1-0328 is
acknowledged.



\begin{references}
\bibitem{landvskubo} Note that derivation of Landauer or Kubo formulas
proceeds from the conceptually different grounds: Landauer
conductance measures a response to the difference in chemical
potentials between two different reservoirs, while Kubo
conductance gives linear DC current as a response to perturbing
the Hamiltonian by a constant electric field. For a physical
situations when this distinction becomes important see
Ref.~\onlinecite{yudson}.
\end{references}
\end{document}